\documentclass[%
 reprint,onecolumn,
 amsmath,amssymb,
 aps,nofootinbib,
]{revtex4-2}
\usepackage{fancyhdr}
\usepackage[colorlinks]{hyperref}
\usepackage[nameinlink,noabbrev]{cleveref}
\hypersetup{
    colorlinks=true,
    linkcolor=blue,
    filecolor=magenta,      
    urlcolor=blue,
   citecolor=blue
}
\usepackage{graphicx}
\usepackage{dcolumn}
\usepackage{bm}

\usepackage{scalerel}
\usepackage{tikz}
\usetikzlibrary{svg.path}

\definecolor{orcidlogocol}{HTML}{A6CE39}
\tikzset{
  orcidlogo/.pic={
    \fill[orcidlogocol] svg{M256,128c0,70.7-57.3,128-128,128C57.3,256,0,198.7,0,128C0,57.3,57.3,0,128,0C198.7,0,256,57.3,256,128z};
    \fill[white] svg{M86.3,186.2H70.9V79.1h15.4v48.4V186.2z}
                 svg{M108.9,79.1h41.6c39.6,0,57,28.3,57,53.6c0,27.5-21.5,53.6-56.8,53.6h-41.8V79.1z M124.3,172.4h24.5c34.9,0,42.9-26.5,42.9-39.7c0-21.5-13.7-39.7-43.7-39.7h-23.7V172.4z}
                 svg{M88.7,56.8c0,5.5-4.5,10.1-10.1,10.1c-5.6,0-10.1-4.6-10.1-10.1c0-5.6,4.5-10.1,10.1-10.1C84.2,46.7,88.7,51.3,88.7,56.8z};
  }
}

\newcommand\orcidicon[1]{\href{https://orcid.org/#1}{\mbox{\scalerel*{
\begin{tikzpicture}[yscale=-1,transform shape]
\pic{orcidlogo};
\end{tikzpicture}
}{|}}}}
\usepackage{amsmath}
\usepackage{amssymb, amsfonts}
\usepackage[greek,english]{babel}
\begin{document}

\preprint{APS/123-QED}

\title{On the collective properties of quantum media}


\author{Kamel Ourabah \orcidicon{0000-0003-0515-6728},}\email{kam.ourabah@gmail.com}
\address{Theoretical Physics Laboratory, Faculty of Physics, University of Bab-Ezzouar, USTHB, Boite Postale 32, El Alia, Algiers 16111, Algeria}

\date{\today}

\begin{abstract}
We discuss the hydrodynamic representation of a wide class of quantum media exhibiting similar elementary excitations and dispersion properties. The representation covers quantum systems characterized by any type of (long-range) self-interaction, associated with an arbitrary potential. It also accounts for possible nonlinearities, which may arise e.g., due to short-range interactions (collisions) in the case of bosons, or from the Pauli exclusion principle for fermions. The approach equally applies to various physical scenarios, such as self-gravitating quantum media (e.g., dark matter), quantum plasmas, Bose-Einstein condensates, and non-condensed cold atomic clouds. We discuss the formal analogies that can be drawn between these different systems and how they can be used to realize laboratory experiments emulating gravitational phenomena, especially in the context of alternative theories of gravity. To substantiate our point, we elaborate more closely on the case of non-minimal matter-curvature coupling gravity theories.
\end{abstract}

\maketitle


\section{Introduction}

Among the four elementary interactions known in Nature, gravity is by far the one that escapes the most laboratory experiments. This is well-known in the regime of general relativity and, to an even larger extent, for quantum gravity. When it comes to studying a large number of self-gravitating particles, this statement holds true even in the Newtonian regime. It is in fact hard to conceive a system of $N$ particles self-interacting solely (or dominantly) through gravity, as competitive interactions are in general orders of magnitude higher than self-gravity. This dramatically restricts experimental studies on the collective properties of self-gravitating systems as well as the study of their relaxation and the corresponding equilibrium profiles. This has motivated the search for \textit{gravity analogs}, i.e., systems exhibiting features similar to gravity, using for example dielectric media, (super)fluids, Bose–Einstein condensates (BECs), etc.
To date, a wide variety of systems has been studied as possible platforms for emulating both Newtonian gravity \cite{NG0,NG1,NG2,NG3} and general relativity phenomena \cite{GR1,GR2,GR3}, and the field of analogs is, by now, a well established one \cite{c1,c2}.

Analogies (from the Greek \textgreek{ανα–λoγoς}, meaning \textit{equal–relation}) translate an underlying invariant structure between two (or more) different objects. In the physical sciences, one speaks of an analogy between two systems when the equations governing them are the same. In these conditions, the formal developments of the theory of one system are valid for the other, provided that due attention is given to some aspects (e.g., the fact that the symbols entering into the equations may have a different meaning or that the boundary conditions for one system may make no sense for the other, or could be very hard to achieve in practice \cite{Alfredo}). This is because “\textit{the same equations have the same solutions}” as Feynmann puts it in his famous “Electrostatic Analogs” lecture \cite{Feynmann}. In practice, one may take advantage of such formal analogies; if one has no access to a certain regime of a target system, but this happens to be equivalent to a reachable regime of an analogous system, one can perform experiments on the latter, and establish results valid for the former.

Bearing these formal analogies in mind, we discuss here a hydrodynamic representation which remains valid in various scenarios of physical interest. The representation is versatile enough to cover many quantum media exhibiting similar elementary excitations, such as \textit{plasmons} in a quantum plasma, \textit{Bogoliubov excitations} in a BEC, and \textit{hybrid phonon modes} in non-condensed cold atomic clouds. The gravitational counterpart of these excitations is the so-called mechanism of Jeans instability (for wave-lengths exceeding a threshold value, known as the Jeans length) and gravitationally-modified sound waves, below this critical value. In the classical limit, the model also applies to the process of chemotaxis, known in biology, which shares many similarities with the process of Jeans instability in self-gravitating media, as discussed e.g. in Refs. \cite{cs,ch2}. 

The paper progresses in the following fashion. In Sec. \ref{II}, we introduce the model. The latter is constructed \textit{such that} it can accommodate any type of interaction and the possible emergence of nonlinearities, which may have various physical origins. In Sec. \ref{III}, we discuss the hydrodynamic representation of this model and derive the corresponding dispersion relation. In Sec. \ref{IV}, we consider specific examples of media where this approach is applicable and examine their dispersion properties. In Sec. \ref{V}, we touch upon a specific class of media (gravity analogs) that can mimic the dispersion properties of non-minimal matter-curvature gravity theory. We conclude in Sec. \ref{VI} with some final remarks. 

\section{The model}\label{II}
To set the scene, let us start our discussion by considering a self-gravitating quantum medium, in the weak-field (Newtonian) regime. In the \textit{mean-field approximation}, such a system is generally described by the Shrödinger-Poisson (also known as the Schrödinger-Newton) model \cite{SN1,SN2,SN3}, namely

\begin{equation}\label{SN}
\begin{aligned}
\mathrm{i} \hbar \frac{\partial \psi}{\partial t} &=\left(-\frac{\hbar^{2}}{2 m} \Delta+m \Phi_{G}\right) \psi, \\
\Delta \Phi_{G} &=4 \pi m G|\psi|^{2},
\end{aligned}
\end{equation}
where $\psi$ is the wave function while $\hbar$, $G$, $m$ and $\Phi_G$ denote, respectively, the reduced Planck constant, the gravitational constant, the mass of the self-gravitating particles, and the gravitational potential. Eq. (\ref{SN}) can equivalently be written in an integro-differential form as follows
\begin{equation}\label{id}
\mathrm{i} h \frac{\partial \psi}{\partial t}=\left[-\frac{\hbar^{2}}{2 m} \Delta-m^{2} G \int \frac{\left|\psi\left(\mathbf{r}^{\prime}, t\right)\right|^{2}}{\left|\mathbf{r}-\mathbf{r}^{\prime}\right|} \mathrm{dr}^{\prime}\right] \psi.
\end{equation}
This is the framework usually adopted for studying quantum matter confined by gravitational fields, as it allows for a simple account of quantum effects in purely gravitational problems. The model (\ref{SN}) finds many applications in astrophysics and cosmology; it describes for instance boson stars \cite{BS1,BS2,BS3} and it is a central ingredient in scalar field dark matter models \cite{SFDM1,SFDM2,SFDM3,SFDM4,Ourabah2020bis,Our2020}.

Interestingly, one may consider a slightly different version of the system (\ref{SN}), applicable to electrons in a quantum (non-magnetized) plasma. In this case, instead of the gravitational potential $\Phi_G$, one has to consider the electrostatic mean-field potential $\Phi_e$, and the counterpart of Eq. (\ref{SN}) reads as
\begin{equation}\label{SP}
\begin{aligned}
\mathrm{i} \hbar \frac{\partial \psi}{\partial t} &=\left(-\frac{\hbar^{2}}{2 m} \Delta-e \Phi_{e}\right) \psi, \\
\Delta \Phi_{e} &=\frac{e}{\varepsilon_{0}}\left(|\psi|^{2}-n_{0}\right),
\end{aligned}
\end{equation}
where $e$ and $\varepsilon_0$ denote, respectively, the elementary charge and the vacuum permittivity, while $n_0$ accounts for the density of the ionic population supposed at rest (i.e., the so-called Jellium model). Eq. (\ref{SP}) is in fact a key element in the description of quantum plasmas (see e.g., \cite{pl0,pl1,pl2}), in particular in studying plasma elementary excitations, i.e., plasmons. Another situation in which Eq. (\ref{SP}) holds is the description of the mean-field behavior of ultra-cold atoms confined
and cooled by laser beams in a magneto-optical trap (MOT). In this case, Eq. (\ref{SP}) formally applies upon defining an effective atomic charge (see e.g., \cite{MOT1,MOT2,MOT3,Oursr})
\begin{equation}\label{Q}
Q \equiv \left(\sigma_{R}-\sigma_{\mathrm{L}}\right) \sigma_{\mathrm{L}} I_{0} / c,
\end{equation}
where $c$ is the speed of light in vacuum, $I_0$ is the intensity of the laser cooling beams, while $\sigma_{R}$ and $\sigma_{L}$ are the radiation and laser absorption cross-sections. In a sense, ultra-cold atoms in MOTs can be regarded as intermediate media, between self-gravitating systems and quantum plasmas, and can mimic the dispersion properties of both, as discussed next. They have recently been considered to build in the laboratory a system of particles with a gravitational-like interaction, using a cold Strontium gas \cite{NG1}. The analogies that can be drawn between these systems, and also with BECs, have been nicely discussed recently by Mendonça through a wave-kinetic approach \cite{Tito}.

As known, Eqs. (\ref{SN}) and (\ref{SP}) can be written in a hydrodynamic form, very useful for studying the elementary excitations taking place in these media. Here, we wish to extend such an approach.
In what follows, we lay out a unified hydrodynamic representation applicable (i) for any type of potential describing self-interactions between the particles and (ii) in the presence of nonlinearities (other than the self-potential itself). Such nonlinearities are known to arise in various situations of physical interest, with different physical origins. One possibility for the emergence of these nonlinearities is the presence of collisions between bosons in a BEC. These collisions can be modeled through a pair contact short-range potential
$u_{coll}\left(\mathbf{r}-\mathbf{r}^{\prime}\right)=g \delta\left(\mathbf{r}-\mathbf{r}^{\prime}\right)$ where $\delta$ is the Dirac delta distribution and $g \equiv 4 \pi a_s \hbar^{2} / m^{3}$ is the coupling constant, $a_s$ being the s-wave scattering length \cite{gpp}. This leads to a nonlinear Schrödinger equation, known as the Gross-Pitaevskii equation \cite{gpp}, 

\begin{equation}\label{gpp1}
i \hbar \frac{\partial \psi}{\partial t}=\left[-\frac{\hbar^{2}}{2 m} \Delta + V+ \frac{4 \pi a_s \hbar^{2}}{m^{3}}  |\psi|^{2}  \right] \psi.
\end{equation}

Another possibility arises in the description of the mean-field behavior of many fermions (e.g., in a quanrum plasma or in fermionic dark matter models), as a consequence of the Pauli exclusion principle. For a nonrelativistic 3d system of spin-$1/2$ particles, this leads to a nonlinear Schrödinger equation of the form\footnote{\textit{Stricto sensu}, Eq. (\ref{gpp2}) is only valid in the 3-dimensional case. In the general $d$-dimensional case, the nonlinearity should read as $\mu(|\psi|^2)= \frac{1}{2}(\frac{d}{2S_d})^{2/d} \frac{(2 \pi \hbar^2)}{m} |\psi|^{4/d}$, $S_d$ being the $d$-dimensional solid angle.} \cite{Manfredi}

\begin{equation}\label{gpp2}
i \hbar \frac{\partial \psi}{\partial t}= \left [-\frac{\hbar^{2}}{2 m} \Delta + V +\frac{\left(3 \pi^{2}\right)^{2 / 3}}{2}  \frac{\hbar^{2}}{m}|\psi|^{4 / 3} \right ] \psi.
\end{equation}
Both Eqs. (\ref{gpp1}) and (\ref{gpp2}) involve a power-law nonlinearity, but this is not necessarily so. Other forms of nonlinearitiy do emerge in diverse physical problems; for example logarithmic nonlinearities, leading to a nonlinear Schrödinger equation of the form\footnote{Nonlinear Schrödinger equations in the form of (\ref{log}) have quite a long history in the mathematical physics literature \cite{log1,log2}. This is because they are the simplest $U(1)$-symmetric wave
equations (apart from the standard linear Schrödinger equation)
which satisfy the dilatation covariance and separability
properties.}
\begin{equation}\label{log}
i \hbar \frac{\partial \psi}{\partial t}=\left[-\frac{\hbar^{2}}{2 m} \Delta+V-b \ln \left({a}|\Psi|^{2}\right) \right] \psi,
\end{equation}
where $a$ and $b$ are constant parameters. Such logarithmic nonlinearities appear in modeling BECs and quantum liquids \cite{log3,log,log4}. They also arise in some dark matter models, due to an effective temperature \cite{logDM1,logDM2,logDM3}. 

In order to account for all these different possibilities, we shall discuss in the following the hydrodynamic representation of the general equation
\begin{equation}\label{1}
i \hbar \frac{\partial \psi}{\partial t}=\left[- \frac{\hbar^2 }{2m}\Delta+V_0 +\int\left|\psi\left(\mathbf{r}^{\prime}\right)\right|^{2} V\left(\left|\mathbf{r}-\mathbf{r}^{\prime}\right|\right) d \mathbf{r}^{\prime} + \mu (\left|\psi\left(\mathbf{r}\right)\right|^{2} ) \right] \psi,
\end{equation}
where $V\left(\left|\mathbf{r}-\mathbf{r}^{\prime}\right|\right)$ represents some generic self-potential and $\mu (\left|\psi\left(\mathbf{r}\right)\right|^{2} )$ is an arbitrary nonlinearity, while $V_0$ denotes a possible external potential (for example due to a neutralizing ionic background in a quantum plasma). Eq. (\ref{1}) covers the aforementioned physical scenarios and other possibly relevant situations as well. Among the potentials covered by Eq. (\ref{1}), those of a Poisson-type are given by
\begin{equation}
V\left(\left|\mathbf{r}-\mathbf{r}^{\prime}\right|\right)=\frac{\mathcal{G}}{\left|\mathrm{r}-\mathrm{r}^{\prime}\right|},
\end{equation}
where (i) $\mathcal{G} \equiv -m^{2} G$ corresponds to a self-gravitating medium \cite{BS1,BS2,BS3,SFDM1,SFDM2,SFDM3,SFDM4,Ourabah2020bis} (see Eq. (\ref{id})), 
(ii) $\mathcal{G} \equiv e^2/4 \pi \epsilon_0$ to electrons in a quantum plasma \cite{pl0,pl1,pl2}, and (iii) $\mathcal{G} \equiv Q/ 4 \pi$ to cold atoms in MOTs \cite{MOT1,MOT2,MOT3}. Another less obvious problem belonging to this class, although only in the classical limit, is the process of chemotactic aggregation describing the collective behavior of microscopic organisms, which can also be modeled by a Poisson-type of potential and which shares many similarities with self-gravitating media \cite{cs,ch2}.


\section{Hydrodynamic representation and dispersion relations} \label{III}

In this section, we present the hydrodynamic representation of Eq. (\ref{1}) and derive a genetic dispersion relation, valid in all the physical scenarios discussed above. For that purpose, we make use of the so-called Madelung (or better Madelung-de Broglie-Bohm) transformation \cite{Madelung}, upon writing the wave function $\psi$ in polar form

\begin{equation}\label{polar}
\psi(\mathbf{r}, t)=  A(\mathbf{r}, t) e^{i S(\mathbf{r}, t) / \hbar},
\end{equation}
where $A(\mathbf{r}, t)$ and $S(\mathbf{r}, t)=(\hbar / 2 i) \ln \left(\psi / \psi^{*}\right)$ are real functions, representing the amplitude and the phase of the wave function. The (number) density and the velocity field are defined in terms of $A$ and $S$ as\footnote{Throughout the paper, the wave function is normalized to the number density $n(\mathbf{r},t)$. In purely gravitational problems, it is standard practice to normalize the wave function to the mass density, i.e., $|\psi(\mathbf{r},t)|^2 \equiv \rho(\mathbf{r},t) = m n(\mathbf{r},t)$.} \cite{Madelung}

\begin{equation}
n(\mathbf{r}, t)=  |\psi|^{2}=   A(\mathbf{r}, t)^2 \quad \text { and } \quad \mathbf{u}=\frac{\nabla S}{m }=\frac{i \hbar}{2 m } \frac{\psi \nabla \psi^{*}-\psi^{*} \nabla \psi}{|\psi|^{2}}.
\end{equation}
Note that, so defined, the velocity field is irrotational, i.e., $\nabla \times \mathbf{u}=\mathbf{0}$. Substituting the wave function (\ref{polar}) into Eq. (\ref{1}) and splitting apart the real and imaginary parts, the imaginary part gives the continuity equation
\begin{equation}\label{con}
\frac{\partial n}{\partial t}+\nabla \cdot(n \mathbf{u})=0,
\end{equation}
while the real part leads to
\begin{equation}\label{8}
\frac{\partial S}{\partial t}+\frac{1}{2 m}(\nabla S)^{2}+V_0+ \int A^2(\mathbf{r'}) V\left(\left|\mathbf{r}-\mathbf{r}^{\prime}\right|\right) d\mathbf{r}^{\prime}+\mu(A(\mathbf{r}))+Q=0,
\end{equation}
where
\begin{equation}
Q \equiv -\frac{\hbar^{2}}{2 m} \frac{\Delta \sqrt{n}}{\sqrt{n}}=-\frac{\hbar^{2}}{4 m}\left[\frac{\Delta n}{n}-\frac{1}{2} \frac{(\nabla n)^{2}}{n^{2}}\right]
\end{equation}
is usually referred to as the quantum potential or the Bohm potential. Eq. (\ref{8}) represents an extension to the Hamilton-Jacobi equation, affected by the quantum potential $Q$, the self-potential $V$, and the nonlinearity $\mu$. Taking the gradient of Eq. (\ref{8}) (and noting that $\nabla \times \mathbf{u}=\mathbf{0}$), one ends up with an Euler-like equation
\begin{equation}
m \left[\frac{\partial \mathbf{u}}{\partial t}+(\mathbf{u} \cdot \nabla) \mathbf{u} \right]=-\nabla V_0-\nabla \left [\int A^2(\mathbf{r'}) V\left(\left|\mathbf{r}-\mathbf{r}^{\prime}\right|\right) d\mathbf{r}^{\prime}  \right ]- \nabla \mu -\ \nabla Q,
\end{equation}
or, equivalently
\begin{equation}\label{Euler}
m \left[\frac{\partial \mathbf{u}}{\partial t}+(\mathbf{u} \cdot \nabla) \mathbf{u} \right]=-\nabla V_0-\nabla \left [\int A^2(\mathbf{r'}) V\left(\left|\mathbf{r}-\mathbf{r}^{\prime}\right|\right) d\mathbf{r}^{\prime}  \right ]- \frac{\nabla p}{n}  - \nabla Q,
\end{equation}
where we have defined the pressure as
\begin{equation}\label{p}
{\nabla p(\mathbf{r})}  = {n(\mathbf{r})} \nabla \mu   (\mathbf{r}).
\end{equation}
Note that $\mu$ being a function of $n:=|\psi|^2$, Eq. (\ref{p}) corresponds to a \textit{barotropic} equation of state, i.e., $p(\mathbf{r})=p(n(\mathbf{r}))$.  

Eqs. (\ref{con}) and (\ref{Euler}) form the hydrodynamic representation of Eq. (\ref{1}). They can be combined in a single equation as follows

\begin{equation}
m \left[ \frac{\partial (n \mathbf{u})}{\partial t} + \nabla (n \mathbf{u} \otimes \mathbf{u}) \right]=- \nabla V_0 - \nabla \left [\int A^2(\mathbf{r'}) V\left(\left|\mathbf{r}-\mathbf{r}^{\prime}\right|\right) d\mathbf{r}^{\prime}  \right ]- \frac{\nabla p}{n}  - \nabla Q.
\end{equation}
To arrive at a generic dispersion relation, we restrict ourselves to the case of small perturbations in a stationary, infinite,
homogeneous, and isotropic equilibrium medium, and write

\begin{equation}
n(\mathbf{r},t)= n_0+ \delta n (\mathbf{r},t) \quad \text{and} \quad  \mathbf{u}(\mathbf{r},t)= \mathbf{u}_0+ \delta \mathbf{u}(\mathbf{r},t),
\end{equation}
where $\delta n$ and $\delta \mathbf{u}$ are supposed small. Upon linearizing the fluid equations (\ref{con}) and (\ref{Euler}), we obtain
\begin{equation}\label{H1}
\begin{aligned}
\frac{\partial \delta n (\mathbf{r},t)}{\partial t}&+n_0 \nabla \cdot \delta \mathbf{u} (\mathbf{r},t) =0,\\
m\frac{\partial \delta \mathbf{u} (\mathbf{r},t)}{\partial t}&=  -\nabla \left [\int \delta n(\mathbf{r'},t) V\left(\left|\mathbf{r}-\mathbf{r}^{\prime}\right|\right) d\mathbf{r}^{\prime}  \right ]-\frac{c_s^2 \nabla \delta n(\mathbf{r},t)}{n_0}- \frac{\hbar^2}{4m}\nabla \cdot (\Delta \delta n (\mathbf{r},t)),
\end{aligned}
\end{equation}
where we have imposed that

\begin{equation}\label{imp}
V_0 =- n_0  \left [\int  V\left(\left|\mathbf{r}-\mathbf{r}^{\prime}\right|\right) d\mathbf{r}^{\prime}  \right ].
\end{equation}
For a quantum plasma, this amounts to imposing that the electrostatic potential satisfies the Poisson equation for the neutralizing ionic background. For a self-gravitating medium, one may ignore the two terms in Eq. (\ref{imp}) altogether, by invoking the so-called “Jeans swindle"\footnote{The Jeans approach of self-gravitating systems is known to suffer from a mathematical inconsistency from the start. This inconsistency resides in that, for a constant density $n_0$, one cannot simultaneously satisfy the Poisson equation $\Delta \Phi = 4 \pi G m n_0$ and the condition of hydrostatic equilibrium $\nabla p + m n \nabla \Phi$, which reduces to
$\nabla \Phi=0$ for a barotropic fluid with a constant density $n_0$. Jeans \cite{Jeans} removed this
inconsistency by assuming that the Poisson equation is sourced \textit{only} by the perturbation and \textit{not} by the background density $n_0$. In our setup, this amounts to ignoring the two members of Eq. (\ref{imp}) altogether. Although this may seem \textit{ad hoc}, it has been shown to be a rigorous mathematical procedure \cite{swindle1,swindle2}. One way around to avoid the “Jeans swindle" is to account for the universe expansion \cite{swindle3}, which introduces a sort of neutralizing background, leading to a modified Poisson equation of the form $\Delta \Phi = 4 \pi G m a(t)^2 (n- n_0)$, in the spirit of the Jellium model of plasma physics ($a(t)$ being the scale factor). Another possibility is to study the dynamical stability of an \textit{inhomogeneous} distribution of matter in a finite domain (box) \cite{swindle4}.}.
Performing a Fourier transform of Eq. (\ref{H1}) and combining the resulting equations, one arrives at the following generic dispersion relation
\begin{equation}\label{gdr}
\omega^{2}=\left(n_{0} / m\right) \tilde{V}(k) k^{2} + c_s^2 k^2+ \frac{\hbar^{2}}{4 m^{2}} k^{4},
\end{equation}
where 
\begin{equation}\label{Vk}
\tilde{V}(k)=\int V(r) \exp [-i \mathbf{k} \cdot \mathbf{r}] d \mathbf{r}
\end{equation}
is the Fourier transform of the self-potential $V$ and 

\begin{equation}\label{cs}
c_{s}^{2}= \frac{1}{m} \left(\frac{d p}{d n} \right)_{n=n_0}
\end{equation}
is the squared speed of sound in the medium, which ultimately stems from the nonlinearity $\mu (\left|\psi\left(\mathbf{r}\right)\right|^{2} ) $ in Eq. (\ref{1}) [see Eq. (\ref{p})]. Let us explicitly work out the pressure term and the associated sound speed for the three main nonlinearities discussed in Sec. \ref{II}. For a BEC, described by the Gross-Pitaevskii equation (\ref{gpp1}), one has

\begin{equation}\label{p25}
p= \frac{2 \pi a_s \hbar^2}{m} n^2 \Longrightarrow   c_{s}^{2}=\frac{4 \pi a_{s} \hbar^{2}}{m^{2}} n_0.
\end{equation}
In this case, $c_s^2$ corresponds to the Bogoliubov squared speed of sound. For fermions, using Eq. (\ref{gpp2}), one has
\begin{equation}\label{p26}
p=\frac{1}{20}\left(\frac{3}{\pi}\right)^{2 / 3} \frac{h^{2}}{m} n^{5 / 3}  \Longrightarrow   c_{s}^{2}=\frac{1}{12}\left(\frac{3}{\pi}\right)^{2 / 3} \frac{h^{2}}{m^{2}} n_0^{2 / 3}.
\end{equation}
In this case, $c_s^2$ is linked to the Fermi velocity $v_F$ through $c_s^2= v_F^2/3=k_B T_F/m$ ($T_F$ being the Fermi temperature and $k_B$ is the Boltzmann constant).
For a logarithmic nonlinearity [viz. Eq. (\ref{log})], one finds

\begin{equation}\label{plog}
p= -b n \Longrightarrow c_s^2 = -\frac{b}{m}.
\end{equation}
In some dark matter models, this logarithmic nonlinearity is associated with an effective temperature, i.e., $b= -k_B T$ \cite{logDM1,logDM2}. In this case, the pressure (\ref{plog}) corresponds to an \textit{isothermal} equation of state. It may be interesting to observe that, for the three types of nonlinearity discussed above, the pressure term corresponds to a special case of a barotropic equation of state, namely a \textit{polytropic} equation of state

\begin{equation}
p \propto  n^{\gamma}, \quad \gamma :=1+\frac{1}{\alpha},
\end{equation}
where $\alpha$ is known as the \textit{polytropic index}. This allows drawing many analogies for example between bosonic and fermionic dark matter models (see e.g., Ref. \cite{Our2022}). Note that, in general, the pressure $p$ (and consequently $c_s^2$) need not be positive, in opposition with the pressure of a normal fluid at finite temperature. A typical example of a negative pressure is that of a BEC with attractive short-range interactions, which is characterized by a negative scattering length $a_s<0$, leading to $c_s^2<0$ \cite{Dalfovo}.

Eq. (\ref{gdr}) is a very general dispersion relation, valid in various scenarios of physical interest. It applies to any kind of self-interaction and it accounts for the possible emergence of nonlinearities. In the next section, we shall discuss some specific examples covered by this dispersion relation.

\section{Specific examples} \label{IV}

We examine in this section some specific examples of the generic dispersion relation (\ref{gdr}).
\subsection{Free particles}

The simplest case is the situation where there is no self-interaction between the particles at all, i.e., $V=0$. In this case Eq. (\ref{gdr}) reduces to

\begin{equation}\label{V00}
\omega^{2}= c_s^2 k^2+ \frac{\hbar^{2}}{4 m^{2}} k^{4}.
\end{equation}
For the particular case where there is no nonlinearity $c_s^2=0$, Eq. (\ref{V00}) simply expresses the kinetic energy, $\epsilon \equiv \hbar \omega = p^2/2m$, with $p = \hbar k$, while in the classical limit ($\hbar \to 0$), one has $\omega^2=c_s^2 k^2$. In this limit, two cases have to be distinguished, namely $c_s^2 >0$ and $c_s^2<0$. In the first case, one has $\omega^2>0$ ($\omega$ real) and the system is \textit{stable} for all $k$; the perturbation oscillates with an angular frequency $\omega = \pm c_s k$, corresponding to a sound wave (a phonon in the language of superfluidity). In the other case ($c_s^2<0$), one has $\omega^2<0$ ($\omega$ imaginary) and the system is \textit{unstable} for all modes $k$; the perturbation evolves exponentially rapidly with time, with a rate $\gamma = \pm (|c_s^2|)^{1/2}k$, which diverges for $k \to \infty$.

When both terms are retained, Eq. (\ref{V00}) applies for example to a BEC without long-range interactions \cite{Chavanisent}. For $c_s^2>0$, the perturbation is stable for all modes and the angular frequency reads as 
\begin{equation}
\omega=\pm \sqrt{\frac{\hbar^{2} k^{4}}{4 m^{2}}+c_{s}^{2} k^{2}}.
\end{equation}
This corresponds to the dispersion relation established by Bogoliubov \cite{Bogoliubov} in order to explain superfluidity, for bosons with a repulsive
self-interaction, i.e., Bogoliubov modes. 
On the contrary, one may consider the case $c_s^2<0$, which may occur for bosons with attractive self-interaction. This limit applies, for example, to axions in the early
universe which have an attractive self-interaction and a negligible self-gravity \cite{Chavanisent} (this leads to the notion of \textit{axitons} in the relativistic regime; see e.g. \cite{Kolb}). In this case, one may see that there is a critical wave-number
\begin{equation}
k^{*}=\left(\frac{4 m^{2}\left|c_{s}^{2}\right|}{\hbar^{2}}\right)^{1 / 2}
\end{equation}
such that for any $k<k^{*}$, the frequency is imaginary ($\omega^2<0$) and the modes are unstable. As there is no (long-range) interaction in this case, this type of instability is a purely hydrodynamical (tachyonic) one. 

Note finally that the crossover between the two limiting cases discussed above ($c_s^2=0$ and $\hbar \to 0$) is determined by the perturbation wave-length. In fact, in the long wave-length limit ($k \to 0$), Eq. (\ref{V00}) predicts $
\omega^2 \sim  c_{s}^2 k^2$ whereas in the short wave-length limit ($k \to \infty$), one has $\omega^2 \sim  \hbar^2 k^{4} / 4 m^2$.

\subsection{Poisson-type self-interaction}

Another interesting class of media is the one corresponding to a Poisson-type self-interaction, i.e., a self-potential inversely proportional to the distance,

\begin{equation}
V\left(\mathbf{r}-\mathbf{r}^{\prime}\right) = \frac{\mathcal{G}}{\left|\mathbf{r}-\mathbf{r}^{\prime}\right|} \quad \Longleftrightarrow \quad \tilde{V}(\mathbf{k}) = \frac{4 \pi \mathcal{G}}{\mathbf{k}^{2}}.
\end{equation}
This scenario covers the case of (i) a self-gravitating medium (for $\mathcal{G} \equiv -m^2 G$), (ii) electron states in a quantum plasma (for $ \mathcal{G} \equiv e^2 / 4 \pi \varepsilon_0$), and (iii) a non-condensed cold atomic cloud (for $\mathcal{G} \equiv Q/4 \pi$). In this case, the generic dispersion relation (\ref{gdr}) reduces to 
\begin{equation}\label{drp}
\omega^{2}= \frac{ 4 \pi  \mathcal{G} n_{0}}{m}   + c_s^2 k^2+ \frac{\hbar^{2}}{4 m^{2}} k^{4}.
\end{equation}
Let us examine more closely some specific examples of Eq. (\ref{drp}). In the case of a self-gravitating medium ($\mathcal{G}=-m^2G$), Eq. (\ref{drp}) reduces to
\begin{equation}\label{drg28}
\omega^{2}= - \Omega_J^2   + c_s^2 k^2+ \frac{\hbar^{2}}{4 m^{2}} k^{4},
\end{equation}
where $\Omega_J \equiv \sqrt{4 \pi G m n_0}$ is Jeans frequency. The dispersion relation (\ref{drg28}) has been studied in the context of dark matter, e.g., by Khlopov \textit{et al.} \cite{Kh} and by Chavanis \cite{SFDM3}. It applies for instance to BEC dark matter (fuzzy dark matter) models, with \cite{Chavanisent} $c_s^2 = 4 \pi a_s \hbar^{2} n_{ 0} / m^{2}$ (cf. Eq. (\ref{p25})). It also applies to fermionic dark matter (e.g., massive neutrinos) for \cite{SFDM3} $c_{s}^{2}=(3 / \pi)^{2 / 3} h^{2}  n_{0}^{2 / 3} / 12 m^{2}$ (cf. Eq. (\ref{p26})), where the nonlinearity arises from the Pauli exclusion principle. This dispersion relation reveals the emergence of unstable modes, i.e., Jeans instability. In fact, one sees from Eq. (\ref{drg28}) that there is a critical wave-number (the Jeans wave-number)
\begin{equation}\label{kj}
k^{*}=\frac{\sqrt{2} m}{\hbar}\left[\sqrt{c_{s}^{4}+\frac{4 \pi G n_0 \hbar^{2}}{m}}-c_{s}^{2}\right]^{1/2} 
\end{equation}
such that, for all modes $k>k^*$, one has $\omega^2>0$, leading to gravitationally-modified waves, while for all modes $k<k^*$, one has $\omega^2<0$; the perturbation evolves exponentially with time with a rate $\gamma = \pm \sqrt{- \omega^2}$. The critical wave-number (\ref{kj}), separating between the two regimes, expresses the interplay between the attractive gravity, the repulsive quantum potential, and
the effect of nonlinearity, which may act as an attractive ($c_s^2<0$) or a repulsive ($c_s^2>0$) self-interaction. The case $c_s^2<0$ opposes to gravity and decreases the critical wave-number $k^*$ while the case $c_s^2>0$ increases the value of $k^*$, allowing for an instability for shorter wave-lengths.

Note that, in the special case where there is no nonlinearity ($c_s^2=0$), the dispersion relation (\ref{drg28}) reduces to

\begin{equation}\label{drg0}
\omega^{2}= - \Omega_J^2   + \frac{\hbar^{2}}{4 m^{2}} k^{4}.
\end{equation}
This case has been studied for example in Refs. \cite{SFDM2,Kh,Bianchi} for bosonic dark matter without short-range interaction. The instability in this case appears for all modes $k<k^*$ with $k^*$ given by
\begin{equation}
k^{*}=\left(\frac{16 \pi G n_0 m^{3}}{\hbar^{2}}\right)^{1 / 4},
\end{equation}
expressing the interplay between gravity and quantum pressure. From another hand, in the classical limit ($\hbar \to 0$), Eq. (\ref{drg28}) reduces to
\begin{equation}\label{drg}
\omega^{2}= - \Omega_J^2   + c_s^2 k^2.
\end{equation}
The situation here depends on the sign of $c_s^2$: For $c_s^2<0$, one always has $\omega^2<0$ and the perturbations are unstable for all modes $k$, while for $c_s^2>0$, the onset of instability is determined by the interplay between the gravitational attraction and the effects of nonlinearity; the instability occurs for all modes $k$ below
\begin{equation}
k^{*}=\left(\frac{4 \pi G n_0 m}{c_{s}^{2}}\right)^{1 / 2}.
\end{equation}
This formally corresponds to the critical wave-number derived by Jeans for a classical barotropic gas \cite{Jeans}. It also applies to BEC dark matter in the so-called Thomas-Fermi approximation \cite{Chavanisent} (where the quantum potential can be neglected). 

Another instance of a Poisson-type potential is the case of electrons in a plasma where one has $ \mathcal{G}=e^2 / 4 \pi \varepsilon_0$. In this case, the dispersion relation (\ref{drp}) reduces to
\begin{equation}\label{drpp}
\omega^{2}=  \Omega_p^2   + c_s^2 k^2+ \frac{\hbar^{2}}{4 m^{2}} k^{4},
\end{equation}
where $\Omega_{p} \equiv \sqrt{e^{2} n_{0} / \varepsilon_{0} m}$ is the plasma angular frequency. The dispersion relation (\ref{drpp}) has been studied by Bohm and Pines \cite{Bohm} in the absence of nonlinearity, and by Ferrel \cite{Ferrel} in the general case, where the nonlinearity arises from the Pauli exclusion principle.

In opposition with the gravitational case, the angular frequency $\omega$ here is real for all modes $k$ and there is no instability. Eq. (\ref{drpp}) shows two dispersion terms: quantum effects introduce a dispersion term proportional to $k^4$ while the nonlinearity induces a dispersion term proportional to $k^2$. 

When the nonlinearity term is ignored ($c_s^2=0$), the dispersion relation (\ref{drpp}) also applies to hybrid phonons \cite{MOT2}, i.e., sound waves taking place in ultra-cold matter in MOTs, provided that $\sigma_{L}<\sigma_{R}$. In fact, in this case, the effective charge $Q$ is positive (see Eq. (\ref{Q})) and one may define an effective plasma frequency,
valid for the neutral gas, as $\Omega_{MOT} \equiv \sqrt{Q n_0  /  m}$; the dispersion relation (\ref{drpp}) remains invariant in this case. For $\sigma_{L}>\sigma_{R}$ however, the quantity $Q$ becomes negative, and $\Omega_{MOT}$ becomes imaginary. In this case, one formally has Eq. (\ref{drg0}) (with $\Omega_J$ replaced by $ \sqrt{ |Q| n_0  /  m}$) and the system behaves as a self-gravitating
medium. In this sense, cold atoms in  MOTs can be thought of as intermediate cases between self-gravitating media and quantum plasmas in terms of their dispersion properties, and can mimic the dispersion properties of both media (see for instance \cite{Tito} for a more elaborate discussion). 

\subsection{Classical limit: Application to chemotaxis}

It may be instructive to examine in further detail the classical limit of the present model and the generic dispersion relation (\ref{gdr}). This limit applies to the physical scenarios discussed above, in their classical regime; e.g., a classical plasma or a BEC in the Thomas-Fermi approximation. Let us discuss here a less obvious situation where the classical limit of this model can be found meaningful, namely the process of chemotaxis.

Chemotaxis is the process through which a population of motile cells (or even social insects) moves in the direction of the higher (attractive) or lower (repulsive) concentration of some chemical agent. Both attractive (positive) and repulsive (negative) chemotaxis are commonplace in life \cite{ch1,plus}; many bacteria (e.g., \textit{Escherichia coli}) move towards the highest concentration of oxygen or glucose in their search for food, while many others go away from unfavorable substances (e.g., alcohols, acids, etc.). Because of its attractive nature, the case of positive chemotaxis leads to the so-called \textit{chemotaxis collapse} which shows many similarities with the process of Jeans instability (gravitational collapse) in self-gravitating media, as discussed in detail for example in \cite{cs,ch2}. Let us briefly discuss here how the present model and the generic dispersion relation (\ref{gdr}) apply in this situation.

The process of chemotaxis can be described through the following fluid model\footnote{In more accurate models, one also accounts for a friction force, measuring the importance of inertial effects, by adding a term $-\xi \mathbf{u}$ to the right hand side of the second equation in (\ref{30}). This however does not modify the critical value of $k^*$ separating between stable and unstable modes \cite{ch2}.}\cite{ch2}
\begin{equation}\label{30}
\begin{gathered}
\frac{\partial \rho}{\partial t}+\nabla \cdot(\rho \mathbf{u})=0, \\
\frac{\partial \mathbf{u}}{\partial t}+(\mathbf{u} \cdot \nabla) \mathbf{u}=-\frac{1}{\rho} \nabla p+\nabla c,
\end{gathered}
\end{equation}
where $\rho \equiv m n$ and the Euler (momentum-balance) equation involves a barotropic pressure $p( \rho)$ accounting for several effects, like anomalous diffusion or the fact that the "particles" do not interpenetrate, while the concentration $c(\mathbf{r})$ satisfies a Poisson-type equation\footnote{This is a simplified model valid in a limit of large diffusivity of the chemical and assuming that the chemicals do not degrade with time \cite{cs,ch2}. In the general case, the chemicals are assumed to diffuse with a coefficient $D_c$ while they are produced by the organisms at a rate $h$ and are degraded at a rate $k$. In this general scenario, Eq. (\ref{c}) has to be replaced by the time-dependent equation $
\frac{\partial c}{\partial t}=D_{c} \Delta c-k c+h \rho
$.}, namely
\begin{equation}\label{c}
 \Delta c=-\lambda(\rho-\bar{\rho}),
 \end{equation}
$\bar{\rho}$ being the average value of the density and $\lambda>0$ ($\lambda<0$) represents attractive (repulsive) chemotaxis. This is equivalent to the classical limit ($\hbar \to 0$) of our hydrodynamic model (Eqs. (\ref{con}) and (\ref{Euler})) with a potential $V=-m c$ of the form
\begin{equation}
V\left(\mathbf{r}-\mathbf{r}^{\prime}\right) = \frac{\lambda m}{4 \pi} \frac{1}{\left|\mathbf{r}-\mathbf{r}^{\prime}\right|}.
\end{equation}
The generic dispersion relation (\ref{gdr}) in this case reduces to
\begin{equation}\label{wc}
\omega^2=-\lambda \bar{\rho}+ c_{s}^{2} k^{2},
\end{equation}
where we have taken the classical limit $\hbar \to 0$. The dispersion relation (\ref{wc}) has been derived for example by Chavanis and Sire \cite{cs,ch2}. In the attractive case ($\lambda>0$), one may observe that Eq. (\ref{wc}) is formally identical to the dispersion relation (\ref{drg}) of a self-gravitating medium in the classical regime (with $\Omega_J \to \sqrt{\lambda \bar{\rho}}$). In this case, there is a critical wave-number
\begin{equation}
k^{*}=\left( \frac{\lambda \bar{\rho} }{c_{s}^{2}}  \right)^{1 / 2}
\end{equation}
such that all modes $k<k^*$ are unstable, while in the repulsive case ($\lambda <0$), $\omega^2$ is positive for all modes $k$ and there is no instability in this case.
\section{Gravity analogs: non-minimal coupling gravity models} \label{V}

With the generic dispersion relation (\ref{gdr}) at hand, it is a simple exercise —at least in principle— to find a medium exhibiting some desired dispersion properties. This is particularly relevant if one aims to conceive a laboratory experiment having the dispersion properties of some alternative theory of gravity. The process of Jeans instability, and the corresponding gravitationally-modified sound waves, have extensively been studied in various alternative theories of gravity \cite{gr1,gr2,Ourclaudio,Claudio}, and the corresponding dispersion relations are known. For a system, to be considered as a gravity analog of a given alternative theory of gravity, its dispersion relation (\ref{gdr}) should formally coincide with that of the given theory. If this is the case, one may think of such a system as an analog of the given theory, regarding its dispersion properties. As an example, let us briefly discuss here the case of theories with a non-minimal coupling between matter and curvature (NMC for short) \cite{Bert}.

The dispersion relation for a self-gravitating medium in the weak-field regime of NMC gravity reads as \cite{Ourclaudio,Claudio}
\begin{equation}\label{omega}
\omega^2=-\frac{(\alpha \gamma-\beta / 2) k^{2}+\gamma / 4}{1+3 \alpha k^{2}} m n_0   + \frac{\hbar^2 k^4}{4m^2},
\end{equation}
where $\alpha$, $\beta$, and $\gamma$ are parameters of the theory. The special case of Newtonian gravity is recovered for $\alpha= \beta=0$ and $\gamma= 16 \pi G$, in which case Eq. (\ref{omega}) reduces to the dispersion relation (\ref{drg0}). For $\alpha=0$, one is in the regime of pure NMC and the dispersion relation (\ref{omega}) reduces to

\begin{equation}\label{omega2}
\omega^2=(- \frac{\gamma}{4}+ \frac{\beta}{2} k^2)  m n_0   + \frac{\hbar^2 k^4}{4m^2}.
\end{equation}
This case is the most physically relevant and also the simplest to analyze. The dispersion relation in this case also corresponds to that of Eddington-inspired Born-Infeld theory \cite{Ahmed}.
Upon identifying Eq. (\ref{omega2}) with the generic dispersion relation (\ref{gdr}), one may see that this is equivalent to an attractive Poisson-like potential and a squared sound speed $c_s^2 \equiv \beta m n_0/2$ (this has been analyzed for example in \cite{Ahmed} where it were observed that such a modification of gravity produces an effective pressure). Alternatively, in the absence of pressure ($c_s^2=0$), Eq. (\ref{omega2}) is reproduced by the generic dispersion relation (\ref{gdr}) for particles self-interacting through a potential of the form

\begin{equation}
V\left(\left|\mathbf{r}-\mathbf{r}^{\prime}\right|\right) = - \frac{A}{\left|\mathbf{r}-\mathbf{r}^{\prime}\right|} + \frac{B}{\left|\mathbf{r}-\mathbf{r}^{\prime}\right|^3},
\end{equation}
with
\begin{equation}
A \equiv \frac{\gamma m^2}{16 \pi} \quad \text{and} \quad B \equiv \frac{\beta m^2}{4 \pi}.
\end{equation}
That is, a superposition of an attractive potential scaling as $1/| \mathbf{r}|$ (mimicking Newtonian gravity) and a potential scaling as $1/| \mathbf{r}|^3$. Attractive potentials in $1/| \mathbf{r}|$, mimicking Newtonian gravity, can be in principle generated in various media, such as cold atomic clouds in MOTs \cite{NG1}, or using attractive capillary interactions in a fluid \cite{NG2}, or using BECs \cite{NG0}. In this latter medium, it is known that for a fixed orientation of the inter-atomic axis with respect to the external
field, one also has an $1/|\mathbf{r}|^3$ variation of
the interaction energy at near-zone separations \cite{NG0}. The emergence of an effective $1/|\mathbf{r}|^3$ potential suggests that BECs could be considered as possible platforms for emulating NMC gravity in future laboratory experiments.

\section{Conclusion} \label{VI}
We have discussed a hydrodynamic representation valid for a wide class of quantum media of physical interest. The representation accounts for any type of (long-range) self-interaction and the possible emergence of nonlinearities, which may arise e.g., due to collisions in a Bose-Einstein condensate or due to the Pauli exclusion principle in the case of fermions. We have discussed different physical scenarios where this approach is applicable, namely the case of free particles and the case of systems characterized by interactions of a Poisson-type, such as self-gravitating media, electrons in a plasma, or cold atoms in magneto-optical traps. We have analyzed the dispersion properties of these media and the analogies that can be drawn between them. Our approach might help realizing laboratory experiments simulating gravitational phenomena. This is of a particular interest in finding laboratory experiments having the same dispersion properties of alternative gravity theories, as illustrated here in the special case of theories with non-minimal coupling between matter and curvature.

This work may open up new prospects for future research. In particular, a couple of possible directions of research can be mentioned. First, it seems worthwhile to extend the present approach to the quantum-relativistic regime. In fact, a (semi)relativistic quantum fluid theory can be constructed in the same fashion, by making use of the Madelung transformation and an expansion of the Dirac Hamiltonian to second
order in $1/c$ \cite{rel}. Note however that this would restrict the class of systems to be considered. In particular, the case of cold atoms has to be discarded as it always belongs to the nonrelativistic regime. Another promising avenue consists in going beyond the linear treatment presented here, and exploring to what extent these analogies hold in the study of nonlinear structures \cite{nonl}, such as solitons, chock waves, vortices, etc.


\end{document}